\documentclass[]{article}
\usepackage{graphicx}
\usepackage{color}
\usepackage{listings}
\usepackage{fullpage}
\usepackage{amsmath}
\usepackage[utf8x]{inputenc}
\usepackage{import}
\usepackage{setspace}
\usepackage{authblk}
\usepackage{hyperref}
\definecolor{lightgray}{gray}{0.5}
\usepackage{setspace}
\usepackage[font={small,sf}, singlelinecheck=false]{caption}
\usepackage[framemethod=tikz]{mdframed}
\usepackage{color,subfigure}
\definecolor{turquoise}{rgb}{0.129,0.4516,0.4194}

\makeatletter
\newenvironment{ProblemSpecBox}[2]{ 
	\protected@edef\@currentlabelname{#2}
	\protected@edef\@currentlabel{#2}
	\begin{mdframed}[
		innerlinewidth=0.5pt,
		innerleftmargin=10pt,
		innerrightmargin=10pt,
		innertopmargin = 10pt,
		innerbottommargin=10pt,
		skipabove=\dimexpr\topsep+\ht\strutbox\relax,
		roundcorner=5pt,
		frametitle={#1},
		frametitlerule=true,
		frametitlerulewidth=1pt]
	}{
	\end{mdframed}
}
\makeatother

\begin{document}
	
\title{The growth and form of knowledge networks by kinesthetic curiosity}

\author[1]{Dale Zhou}
\author[2]{David M. Lydon-Staley}
\author[3]{Perry Zurn}
\author[2,4,5,6,7,8,9]{Danielle S. Bassett}

\affil[1]{Neuroscience Graduate Group, Perelman School of Medicine, University of Pennsylvania, Philadelphia, PA 19104, USA}
\affil[2]{Department of Bioengineering, School of Engineering and Applied Sciences, University of Pennsylvania}
\affil[3]{Department of Philosophy, American University, Washington, D.C.}
\affil[4]{Department of Physics \& Astronomy, College of Arts and Sciences, University of Pennsylvania}
\affil[5]{Department of Psychiatry, Perelman School of Medicine, University of Pennsylvania}
\affil[6]{Department of Neurology, Perelman School of Medicine, University of Pennsylvania}
\affil[7]{Department of Electrical \& Systems Engineering, School of Engineering and Applied Sciences, University of Pennsylvania}
\affil[8]{Santa Fe Institute, Santa Fe, NM 87501 USA}
\affil[9]{To whom correspondence should be addressed: dsb@seas.upenn.edu}

\maketitle

\begin{abstract}
	 Throughout life, we might seek a calling, companions, skills, entertainment, truth, self-knowledge, beauty, and edification. The practice of curiosity can be viewed as an extended and open-ended search for valuable information with hidden identity and location in a complex space of interconnected information. Despite its importance, curiosity has been challenging to computationally model because the practice of curiosity often flourishes without specific goals, external reward, or immediate feedback. Here, we show how network science, statistical physics, and philosophy can be integrated into an approach that coheres with and expands the psychological taxonomies of specific-diversive and perceptual-epistemic curiosity. Using this interdisciplinary approach, we distill functional modes of curious information seeking as searching movements in information space. The kinesthetic model of curiosity offers a vibrant counterpart to the deliberative predictions of model-based reinforcement learning. In doing so, this model unearths new computational opportunities for identifying what makes curiosity curious.
\end{abstract} 



\newpage

    Seeking information with the potential value to learn diverse skills, understand the world, form social relations, and promote individual well-being is essential to flourishing throughout life \cite{litman2008interest, kidd2015psychology, gottlieb2018towards, kashdan2018five, lydon2019within, wade2019role, GRUBER20191014}. Humans encounter information in an ever-expanding and shape-shifting search space of knowledge that is vast and complex \cite{sole2010language, zurn2019philosophy}. The stream of encounters can bring about averse states of uncertainty, pleasurable states of interest, or expectations of usefulness for learning and action, thereby guiding future exploratory strategies \cite{litman2008interest, kashdan2018five}. However, inferring the hedonic or utilitarian value of information to guide behavior is costly and time-consuming in complex spaces \cite{GOTTLIEB2013585, bossaerts2017computational,  gottlieb2018towards}. Curiosity may have evolved to overcome these costs, promoting efficient search strategies to encounter potentially valuable information without prior knowledge of the information's identity and location \cite{de2011levy, hills2015exploration, gottlieb2018towards}. \\

\begin{figure}[htbp!]
    	\centering
    	\includegraphics[width=1\columnwidth]{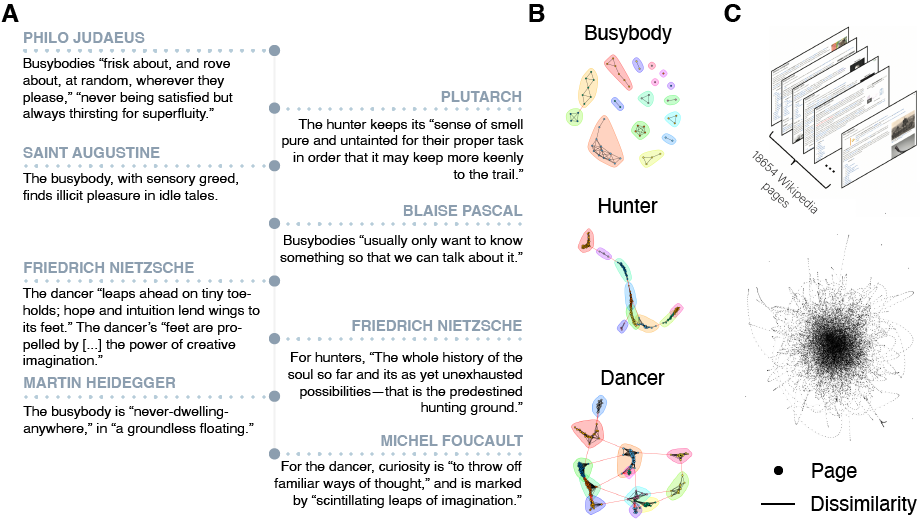}
    	\caption{\textbf{The kinesthetic curiosity model: the concept, associated network representations, and an operationalization in an ecological experiment.} \textbf{\emph{(A)}} The kinesthetic curiosity model posits that curiosity is best explained by movement through information space. Three such movements are conceptually operationalized by the historical archetypes of the \textit{busybody}, the \textit{hunter}, and the \textit{dancer} \cite{zurn2019chapter}. Selected quotes from the 1st to 20th centuries describe each of the modes \cite{yonge1854works, augustine1961confessions, pascal1968pensees, moralia2005, heidegger1996being, foucault2019ethics, nietzsche1996philosophy}. \textbf{\emph{(B)}} The information-seeking movements of the kinesthetic curiosity model can be formalized in the abstract. Each movement is a walk on an underlying collective (or otherwise \emph{a priori} existing) knowledge network. As each person takes this walk, they build their own individualized knowledge network composed of informational units (nodes) and informational relations (edges) \cite{zurn2018curiosity, bassett2020network}. Busybodies construct loose networks, hunters build tight networks, and dancers bridge seemingly disparate modules \cite{zurn2018curiosity, zurn2019chapter}. Here we show simulated knowledge networks that we generated using a computational model of network growth (see main text) \cite{lydon2019hunters}. Modules are colored according to the WalkTrap algorithm \cite{pons2005computing}. \textbf{\emph{(C)}} To make the kinesthetic curiosity model concrete to the reader, we consider the example information network of Wikipedia, whose 5.8 million nodes are articles and whose edges are article-to-article hyperlinks. As humans browse, they walk from article to article by hyperlinks, the search bar, or the random page generator; the distance traversed by the walk reflects the similarity of word usage between documents. The sequence of steps can be used to test and validate the kinesthetic curiosity model of knowledge network growth \cite{lydon2019hunters}. Individual differences in the sequence reflect a unique architectural style of information seeking \cite{lydon2019hunters}: knowledge networks with more hunter-like dynamics (versus busybody-like) were associated with higher sensitivity to uncertainty, lower enjoyment of exploration, and lower sensation seeking.}
    	\label{fig1}
	\end{figure}

    We propose to understand the kinesthetic modes of search strategies associated with curiosity using network science, statistical physics, and philosophy \cite{zurn2018curiosity, zurn2019chapter, lydon2019hunters, bassett2020network}. Since antiquity, investigations of curiosity have contemplated its essential components \cite{zurn2019chapter}. However, we argue that curiosity is best characterized by its searching function \cite{zurn2019chapter}. In a model called \textit{kinesthetic curiosity}, we distill three major modes of potentially many functions described by movement embedded within information landscapes: the \textit{busybody} scouts for loose threads of novelty, the \textit{hunter} pursues specific answers in a projectile path, and the \textit{dancer} leaps in creative breaks with tradition \cite{zurn2019chapter}. Each mode of function is linked with a distinct signature of searching movement (Figure \ref{fig1}). The paths of movement from one piece of information to another are threads creating webs of interlinked information, which we call \textit{knowledge networks}. As the byproduct of searching movement, knowledge network structures may support learning, creativity, and social behavior, without the need for task-specific goals, utility, and feedback \cite{hills2012optimal, abbott2015random, muentener2018efficiency, wheatley2019beyond, kenett2019semantic, lynn2019human, siew2019using, siew2020applications}.  \\
	
    In this review, we first describe the kinesthetic curiosity model of efficient search (Section \ref{section1}). We then explain how kinesthetic curiosity integrates existing theories of curiosity and reinforcement learning (Section \ref{section2}). Next, we consider the evolutionary origins of kinesthetic curiosity and hypothesize neural mechanisms (Section \ref{section3}). Last, we propose that analyses of knowledge network structure and growth quantify previously qualitative descriptions of curiosity and test contemporary theories of information seeking (Section \ref{section4}). Broadly, we offer a perspective that expands our understanding of the practice of curiosity beyond states and traits by positing a computational model of kinesthetic curiosity.
  	
\section{Knowledge network growth principles}\label{section1}
  	
The kinesthetic curiosity model posits open-ended and intrinsically motivated movements in an information space (Figure \ref{fig1}). What is the elementary rule of such movements? To answer this question, we observe that humans often behave quickly and automatically when they must consider many options in complex and changing environments \cite{bossaerts2017computational}. Thus, a natural rule to consider is that of a random walk. In fact, random walk models can explain several intelligent behaviors including exploration, navigation, memory recall, creativity, and foraging \cite{hills2012optimal, abbott2015random, kenett2019semantic}. Random exploration is economical in that it requires little computational capacity, but can oversample the environment and thus be inefficient \cite{findling2019computational, stella2019hippocampal, GOTTLIEB2013585}.\\
 
The random walk model therefore requires additional principles that account for the kinesthetic signatures of discovery and search \cite{berlyne1954theory, tria2015dynamics, PhysRevLett.120.048301, viswanathan1999optimizing}. Two parsimonious movement principles generate the kinesthetic signatures of the busybody, hunter, and dancer, as well as the diverse spectra in between \cite{zurn2019chapter, lydon2019hunters}. First, movements to discover new information can be biased by memory of the familiar; second, search patterns can be biased to efficiently explore a space. The principle of memory for the familiar can be formalized by the notion of \textit{edge reinforcement}, and that of efficient search can be formalized by the notion of a \textit{L\'evy flight}. In isolating these principles of discovery and search, it becomes possible to model individual differences in curious practice \cite{lydon2019hunters}.\\
  	
\subsection{Principle 1: Edge reinforcement}
How do humans discover knowledge \cite{PhysRevLett.120.048301}? Put simply, people learn and innovate by revisiting remnants of the past with a fresh perspective. New flickering patterns emerge from well-trodden hubs that shape the flow of information seeking into new directions, expanding the known unknown to explore newly adjacent possibilities \cite{tria2015dynamics, PhysRevLett.120.048301}.\\

In the kinesthetic curiosity model, random walkers individually vary in their preference for either new or familiar information by the mechanism of \textit{edge reinforcement} \cite{PhysRevLett.120.048301}. Edge reinforcement is a memory of familiarity that increases the probability of taking previously traversed paths. As a cognitive process, edge reinforcement is supported by accurate and rapid learning for a vast visual and social memory of familiarity \cite{Brady2008VisualLM, parkinson2017spontaneous}. As a movement principle, edge reinforcement resembles the recurrent dynamics of many other existing models for exploration and foraging \cite{da2015and, abbott2015random, Bellemare2016UnifyingCE, Jaegle2019VisualNC}. When applied to human information seeking, edge reinforcement is associated with the personality trait of \textit{deprivation sensitivity}, a dimension of curiosity associated with aversion to uncertainty and gaps of knowledge \cite{kashdan2018five, litman2008interest, lydon2019hunters}.\\

Seeking information until a knowledge gap is filled characterizes a persistent and effortful form of specific exploration that resolves an unknown by incorporating new information into existing knowledge \cite{litman2008interest, gweon2014sins, leonard2017infants}. Hunter-like individuals have high deprivation curiosity, creating tighter knowledge networks with greater edge reinforcement as they encounter new information, recognize gaps in their knowledge, and revisit concepts in an iterative cycle of filling in knowledge gaps \cite{loewenstein1994psychology, shin2019homo, lydon2019hunters}.
  	
\subsection{Principle 2: L\'evy flight}
A pervasive scarcity of resources induces organisms to efficiently search for value despite lacking prior knowledge of the search space, location of targets, and identity of targets \cite{viswanathan1999optimizing, wosniack2015robustness, wosniack2015efficient, da2015and, reynolds2018current}. Yet, potential value may be unpredictable across the lifespan, between unique individuals, and in high-dimensional environments \cite{sutton2018reinforcement, salganik2020measuring}. Therefore, we assume that potentially valuable information is sparsely and randomly (unpredictably) distributed in a complex, unknown environment. \\

In these environmental conditions, long-term search efficiency across the lifespan and evolution is often modeled with respect to energetic cost \cite{bartumeus2007levy}. Efficiency is the ratio of resource encounters to energy expenditure, operationalized by the number of steps taken in both spatial and abstract landscapes and the total distance traversed \cite{viswanathan1999optimizing, Bellmundeaat6766, garvert2017map, solomon2019hippocampal}. Consequently, efficiency depends upon the distribution of step distances. The optimal distribution of step distances is thought to be a power-law \cite{viswanathan1999optimizing}, which L\'evy flights produce in their fractal movement patterns characterized by many small steps and a few large steps \cite{viswanathan1999optimizing, wosniack2015robustness, wosniack2015efficient, da2015and}. While L\'evy flights have been frequently studied in environments where reward is sparse and randomly distributed, the dynamics are theoretically optimal in random search across a variety of environmental conditions \cite{viswanathan1999optimizing, da2015and}. However, especially in smaller spaces or shorter timescales, other qualities of search, such as speed, reliability, and robustness, may prove more relevant for diverse individuals and walks of life \cite{bartumeus2007levy, lomholt2008levy, palyulin2014levy}. With respect to other search qualities, the least costly paths are not necessarily the most worthwhile \cite{bartumeus2007levy, leonard2017infants, inzlicht2018effort}. Therefore, long-term inefficiency does not imply individual deficiency. \\
  	
In the kinesthetic curiosity model, a sequence of steps in the random walk weaves a thread through the underlying network. The distance is defined as the number of edges traversed between semantic units \cite{kenett2017semantic}. To assess the existence and extent of L\'evy flight dynamics, we consider the manner in which the probability of a step decays with semantic distance, and we measure the exponent of that decaying probability distribution. Optimally efficient L\'evy flight dynamics exist if this exponent is approximately 2 \cite{viswanathan1999optimizing}. Although the scope and prevalence of L\'evy flight dynamics across organisms remains actively debated \cite{palyulin2014levy, reynolds2018current}, recent work shows that humans indeed exhibit an average exponent of $2.11 \pm 0.15$ suggestive of L\'evy-like dynamics in curiosity-driven information seeking \cite{lydon2019hunters}. Efficiently searching space could explain how individuals acquire a large repertoire of diverse information with limited resources, while avoiding information that is too difficult or too easy to learn \cite{berlyne1954theory, GOTTLIEB2013585, muentener2018efficiency,gottlieb2018towards}.

\section{Integrating models of curiosity and learning}\label{section2}

	\begin{figure}[htbp!]
  		\centering
  		\includegraphics[width=1\columnwidth]{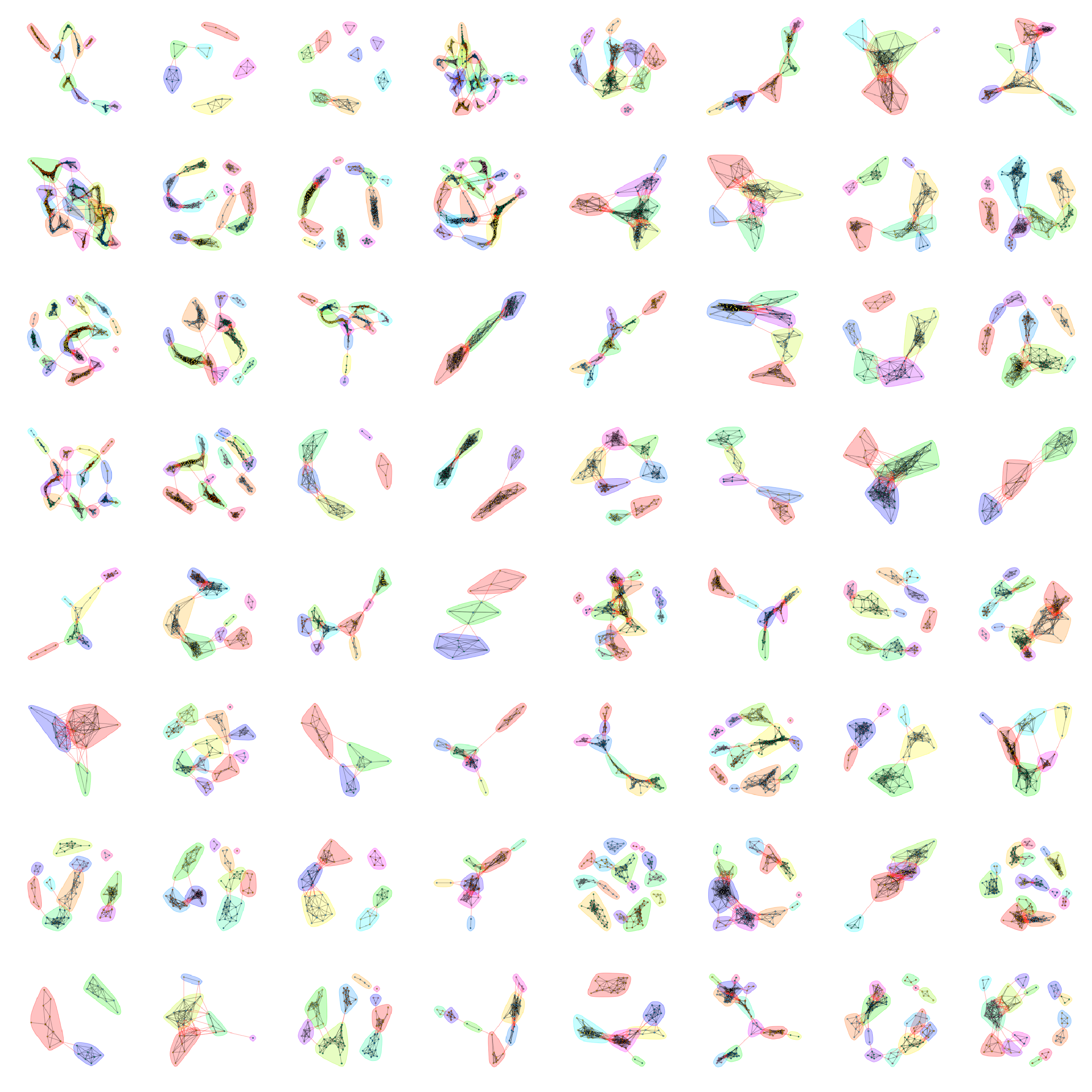}
  		\caption{\textbf{Knowledge network growth, form, and individual variation.} A few simple principles generate diverse kinesthetic signatures of curiosity. Here we show knowledge networks that were randomly generated using the kinesthetic curiosity model encoded as a random walk biased by principles of edge reinforcement and L\'evy flight. Modules are colored by the WalkTrap algorithm \cite{pons2005computing}, and edges connecting modules are colored red. The variation evident in these graphs emphasizes the flexibility of the model to fit individual differences. Differing kinesthetic modes may produce dynamics that grow knowledge networks to be efficiently compressible by exploiting the network structure of hubs and modules \cite{rosvall2008maps, momennejad2017successor, lynn2019human, lydon2019hunters}. Kinesthetic curiosity emphasizes ecological modes of function. Ecologically relevant behaviors are those that individuals are expected to perform in interaction with their daily environments or with similar complexity \cite{falk2013representative}. Greater ecological relevance instills confidence that theories derived from the carefully designed self-report measures and laboratory tasks generalize to some scope of real-world contexts. It equips researchers to study the evolving practice of curiosity in differing contexts throughout recorded history \cite{hills2015exploration, thompson2018lvy,zurn2019chapter,  zurn2019philosophy}. Future work can address the cross-cultural limitations of psychological assessments and models which over-sample homogeneous demographics \cite{falk2013representative}. In practice, the ability to study uncontrived tasks may prove useful for studies in pediatric or clinical study samples, where tasks that are too difficult or unengaging can introduce statistical biases.} 
  		\label{fig2}
  	\end{figure} 

In this section, we posit that the movement principles of kinesthetic curiosity grow knowledge networks that naturally become cognitive maps. Cognitive maps are internal models learned from the relational structure of a stream of experience \cite{tolman1948cognitive}. Hierarchically abstracting the structure of cognitive maps at coarser to finer levels can help people to plan, act, and generalize experiences to novel situations \cite{Bellmundeaat6766, collins2019reinforcement,  momennejad2020learning}. By using information theory to assess the cost of constructing cognitive maps and of abstracting hierarchies, we propose an integration of curiosity and reinforcement learning.

\subsection{Map-taking and map-making}
The predictive processing approaches, including intrinsically motivated reinforcement learning, propose learning predictive models of the world to flexibly guide optimal actions and further learning \cite{GOTTLIEB2013585, stachenfeld2017hippocampus, gottlieb2018towards, collins2019reinforcement, momennejad2020learning}. The predictive models are cognitive maps incorporating prior knowledge of relevant environmental properties \cite{doi:10.1111/tops.12138, Friston2017ActiveIC, gottlieb2018towards, wade2019role}. However, advance knowledge of the environment, including the identity of relevant properties, is often inaccessible. \\

In unknown environments, inferring the value of potential actions becomes computationally prohibitive due to inefficient scaling with the number of explorable units \cite{momennejad2017successor, momennejad2020learning}. This limitation prompts us to consider curiosity's function of efficient search \cite{berlyne1954theory, wosniack2015robustness, wosniack2015efficient, da2015and, reynolds2018current, gottlieb2018towards}. As the search unfolds, information seekers build knowledge networks that naturally become maps due to the emergence of structure from the myriad relationships inherent to sets of semantic units \cite{garvert2017map, solomon2019hippocampal} (Figure \ref{fig2}).
  	 
\subsection{Efficient search enhances the learnability of cognitive maps}
  	 
The cognitive maps arising from kinesthetic curiosity might improve their learnability, conveying information efficiently by omitting unnecessary detail \cite{rosvall2008maps, lynn2019human, zurn2020network, VanKesteren2020}. Recent work reported that to efficiently convey information, network representations of that information should be characterized by highly connected hubs and tightly linked modules \cite{lynn2019human, chai2020evolution}. Hubs and modules are features of hierarchical organization exhibited by diverse signatures of kinesthetic curiosity (Figure \ref{fig2}). Despite individual differences in edge reinforcement and L\'evy flight dynamics, modularity is a core feature of knowledge networks \cite{lydon2019hunters}. \\

People who experience a random sequence of sensory units from a network can learn the network's emergent structure \cite{karuza2016local, schapiro2016statistical, garvert2017map,karuza2017process,Lynn2018StructureFN,kahn2018network, lynn2019human,lynn2019humans}. Similarly, current models of learning and decision-making, such as the successor representation and model-based reinforcement learning, propose to predict, plan, and generalize by hierarchically abstracting structure from sequences of experience \cite{collins2017cost, momennejad2017successor, collins2019reinforcement,  momennejad2020learning}. A sequence of experiences can be direct, such as in information seeking; or it can be indirect, as in the memory replay of recollected experiences \cite{momennejad2020learning}. To flexibly abstract hierarchical structure from a direct or indirect sequence of experiences, individuals would benefit from cognitive maps that are diversely yet economically organized with hubs and modules \cite{collins2017cost, collins2019reinforcement, Bellmundeaat6766}.  
  	 
\subsection{Compression progress theory integrates curiosity and learning}
  	 
Let us now consider how to operationalize the learnability of cognitive map structure. We begin by noting that kinesthetic curiosity is consistent with the \textit{compression progress theory} of curiosity but differs from the \textit{learning progress hypothesis} \cite{schmidhuber2008driven, gottlieb2018towards}. Compression progress theory posits that individuals practice curiosity to seek information that improves compression of their mental model of the world or a sector thereof. Information compression balances the compactness and accuracy of representations, possibly by exploiting redundancy of knowledge network connections in hubs and modules \cite{rosvall2008maps, karuza2019human, lynn2019human}. Whereas compression progress prioritizes compactness, learning progress prioritizes accuracy \cite{schmidhuber2008driven, gottlieb2018towards}. The compression or abstraction of information in many models of learning, memory, and decision-making is achieved by discounting potentially irrelevant information that is more distant in space or time \cite{howard2002distributed, Friston2017ActiveIC, momennejad2017successor, stachenfeld2017hippocampus, lynn2019human}. \\

We hypothesize that kinesthetic curiosity increases learnability by producing knowledge network growth with increasing compressibility. Compressibility is operationalized as the information theoretic codelength (bits per step) required to minimally encode random walks in the knowledge network \cite{rosvall2008maps}. Recalling that humans and reinforcement learning algorithms abstract structure from random sequences of experience, the codelength of a random walk in the knowledge network can be used to measure the cost, or learnability \cite{rosvall2008maps}. A decrease in codelength corresponds to a compression gain. Note, however, that coarser abstractions do not always entail compression gains, because inaccurate coarseness demands overly frequent usage of fewer abstracted components in the encoding. \\

We propose using the operationalization of compressibility as reduced codelength to make three predictions testing the link between curiosity and learning. First, if kinesthetic curiosity is linked with compression progress theory, then random walks biased by the two movement principles of edge reinforcement and L\'evy flight will grow knowledge networks with increasing compressibility. Second, if kinesthetic curiosity is linked with the learnability of cognitive maps, then the strength of hubs and modules in knowledge networks will be linked to compressibility. Third, if kinesthetic curiosity is linked with hierarchical abstraction, then the integrated compressibility that corresponds to finer and coarser abstractions of the same knowledge network will explain how the cost of information motivates discounts according to distance and time. Together, the frameworks of kinesthetic curiosity and predictive processing can be bridged by examining whether knowledge networks are built to be increasingly compressible \cite{rosvall2008maps, momennejad2017successor, stachenfeld2017hippocampus, siew2019using, siew2020applications, lynn2019human, zhou2020efficient}.

\section{Neural implementation and evolutionary origins}\label{section3}

Here, we consider kinesthetic curiosity at the levels of evolution, as well as the micro-, meso-, and macro-scale brain network. We begin by noting that the intrinsic motivation of curiosity purportedly evolved for long-term learning despite rapidly growing complexity in the surrounding habitat \cite{GOTTLIEB2013585, kidd2015psychology}. It is therefore critical to assess how models of curiosity can contend with ecologically relevant complexity. For animals and organisms with extensive limitations on computational capacity, it is challenging to trace the evolutionary history of uncertainty monitoring \cite{bossaerts2017computational, carruthers2017epistemic}. In contrast, the dynamics of kinesthetic curiosity have been observed in the foraging movements of organisms and animals, perhaps evolving for a need to navigate habitats with increasing complexity \cite{viswanathan1999optimizing, de2011levy, sims2014hierarchical, hills2015exploration, wosniack2015robustness, wosniack2015efficient, da2015and, wosniack2017evolutionary, reynolds2018current}. \\

To modulate foraging behaviors at the micro-scale, the neural mechanisms of kinesthetic curiosity likely involve dopaminergic function \cite{hills2015exploration, addicott2017primer}. Consistent with this proposition, prior curiosity research has reported the involvement of dopaminergic brain areas associated with reward anticipation and subjective value, as well as dopaminergic plasticity of the hippocampus in reward-driven associative learning \cite{kang2009wick, GRUBER20191014}. Dopaminergic function for foraging predates function linked to reward anticipation and learning, suggestive of a potential dual role of dopamine in reinforcement learning and kinesthetic curiosity \cite{hills2015exploration, chiew2018motivational, wade2019role}. \\
  	
At the meso-scale, and in contrast to prior curiosity research, we hypothesize a central and concerted role of the hippocampal-entorhinal circuit in curiosity due to the mechanisms that underpin foraging-oriented locomotion, cognitive maps of space, structure learning, and navigation \cite{tolman1948cognitive, schapiro2016statistical, stachenfeld2017hippocampus, garvert2017map, tompson2020functional, Constantinescu1464, Bellmundeaat6766, Mok2019ANA, schuck2019sequential, wade2019role}. Recent neural and behavioral work has reexamined errors and noise due to limited computational capacity as advantageous features of exploration and learning \cite{bartumeus2007levy, mattar2018prioritized, lynn2019human, findling2019computational, stella2019hippocampal, schuck2019sequential}. Similarly, we posit that the limited capacity to optimally choose from potential movement plans for foraging during the hippocampal replay of model-based reinforcement learning results in random movement \cite{mattar2018prioritized, stella2019hippocampal, schuck2019sequential}. L\'evy flight dynamics may emerge from the interaction between individuals with limited cognitive capacity and complex environments \cite{bartumeus2007levy, de2011levy, sims2014hierarchical, stachenfeld2017hippocampus, wosniack2017evolutionary, reynolds2018current}. \\

Last, we predict that macro-scale brain network structure and function moderate individual differences in learning related to the compressibility of knowledge networks \cite{schmidhuber2008driven, lynn2019human, siew2019using, siew2020applications}. We hypothesize the involvement of network hubs in the frontoparietal circuit, which are thought to support executive function, learning, and information compression, as well as the default-mode network associated with mind wandering and hierarchical abstraction \cite{momennejad2017successor, sormaz2018default, summerfield2020structure, mack2020ventromedial, zhou2020efficient}. A set of brain regions across both networks are associated with the reinforcement learning of implicit and explicit representations of experience \cite{dohmatob2018dark, schuck2019sequential}. Together, the neural circuitry shared by model-based reinforcement learning and kinesthetic curiosity suggests that the models are functional counterparts, enacting more learnable experiences for predicting value.

\section{Expanding current taxonomies of curiosity}\label{section4}

An influential psychological taxonomy of curiosity describes the personality trait along the two axes of specific-diversive and perceptual-epistemic \cite{berlyne1954theory, litman2008interest, kidd2015psychology, kashdan2018five}. Here we show how the movement principles and modes of kinesthetic curiosity can quantify and expand these current qualitative frameworks. We also discuss the implications of the kinesthetic curiosity model for understanding the relations between creativity, learning, and social behavior.
		
\subsection{Specific-diversive and perceptual-epistemic curiosity} 
  		
In kinesthetic curiosity, interdigitations of the specific-diversive and perceptual-epistemic axes can be quantified according to movement dynamics producing tight or loose knowledge networks \cite{lydon2019hunters}. Specific curiosity is the desire for particular relevant pieces of information, while diversive curiosity is a general drive to explore different information. These qualitative descriptions underscore the fact that novelty, diversity, and unexpectedness are functions of the sequence or dynamics of information seeking, rather than properties of each element of information \cite{tria2015dynamics, PhysRevLett.120.048301}. In contrast, the second classical axis of curiosity describes the contents of each element of information with perceptual to epistemic properties \cite{berlyne1954theory}. Perceptual curiosity is a drive to seek novel sensory information, whereas epistemic curiosity is the drive to learn conceptual knowledge and regulate uncertainty \cite{kidd2015psychology}. Recent work applied kinesthetic curiosity to quantify the specific-diversive axis of epistemic curiosity \cite{lydon2019hunters}. People who sought specific information during Wikipedia browsing constructed knowledge networks that were tighter than those who sought diverse information. Future tests of perceptual curiosity could apply kinesthetic curiosity to assess how people seek images and videos \cite{heusser2018experience, Jaegle2019VisualNC}. \\
 		
Contemporary theories of epistemic curiosity state that information seeking is governed by a desire to close an information gap between current uncertainty and preferred baseline uncertainty \cite{loewenstein1994psychology}. However, it remains unclear how to define an individual's preference for uncertainty \cite{kidd2015psychology, GOTTLIEB2013585}. The information gap theory addresses the question of \textit{what} one needs to know, whereas kinesthetic curiosity addresses the sense \textit{that} one needs to know. The latter allows researchers to investigate the growth dynamics that arise from extended preferences. Prior work has reported that individuals with a stronger personality trait of deprivation sensitivity, a preference for certainty, produced tighter knowledge networks with closed cycles and greater edge-reinforcement \cite{kashdan2018five, lydon2019hunters}. Therefore, the preferred uncertainty is linked to the deprivation sensitivity dimension of trait curiosity. A future test of the information gap theory could assess how modes of movement fill topological gaps in the knowledge network, as children fill gaps in semantic networks when learning language or explore when information is incomplete \cite{gweon2014sins, Sizemore2017KnowledgeGI}.
 		
\subsection{The dancer, creativity, learning, and social behavior}
In addition to the tracking hunter and collecting busybody, the dancer is a third historical archetype of the practice of curiosity \cite{zurn2019chapter}. In the kinesthetic curiosity model, it seems natural to posit that the dancer is characterized by L\'evy flight dynamics with optimal efficiency. The dancer lacks a direct analog to the trait curiosity self-report measures and classical psychological axes of specific-diversive and perceptual-epistemic curiosity \cite{lydon2019hunters, kashdan2018five, kidd2015psychology}. Yet, due to its emphasis on linking prior knowledge with leaps to new unknowns, the dancer may be the mode of curiosity that best explains the links of curiosity with creativity, learning, and social behavior \cite{magid2015imagination, falk2018persuasion, wheatley2019beyond, Gray2019ForwardFA, zurn2019chapter, kenett2019semantic}. A highly creative person's semantic network displays robustness and hierarchical organization, as does efficient communication \cite{kenett2019semantic, wheatley2019beyond, lynn2019human, chai2020evolution}. Thus, it would seem natural in future studies of curiosity, creativity, and social behavior to determine whether the dancer, moreso than other modes, builds knowledge networks with greater robustness and more hierarchical organization. 

\section{Conclusion}
 	
The kinesthetic curiosity model can formalize, assess, and expand the classical psychological taxonomy of curiosity. We present a computational model of curiosity that hues close to its ecological function in naturalistic environments and to its need to efficiently contend with complexity. The implicit construction of learnable knowledge networks accompanies the explicit learning of structure, as random search may arise from the neural capacity limits for learning. The costs of abstracting hierarchical structure from cognitive maps could link curiosity, creativity, social behavior, and learning under the framework of compression progress theory. In our view, the practice of curiosity includes distinct modes of information seeking dynamics that package information into knowledge networks with unique structural signatures. Considering the philosophical archetypes of curiosity embodied in the hunter, busybody, and dancer will help us to understand curiosity throughout history, across cultures, and at the scales of individuals and societies.

\newpage
 		\begin{ProblemSpecBox}{Box 1. Outstanding questions}{Box 2}
 			\label{box2}
 			
		    How does knowledge network structure and information compression influence learning, creativity, and social interactions? \\
			
 			How could the brain implement kinesthetic curiosity with the mechanisms of foraging, spatial navigation, cognitive maps, and information compression?\\
 			
 			How does kinesthetic curiosity differ with individual variation of attention, mood, motivation, learning, and social behavior, and in psychiatric disorders?\\
 			
 			How should institutions of science, education, media, and markets create incentives for differing modes of kinesthetic curiosity?
 		\end{ProblemSpecBox}
\newpage

\section{Highlighted references}
\begin{itemize} 
	\item P. Zurn, “Chapter two busybody, hunter, dancer: Three historical models of curiosity,” Toward New Philosophical Explorations of the Epistemic Desire to Know: Just Curious About Curiosity, p. 26, 2019.
	\begin{itemize} 
		\item This philosophy paper proposes the kinesthetic model of curiosity, identifying three modes of curious movement: the busybody, hunter, and dancer. To this end, the paper develops a conceptual framework to understand the practice of curiosity and its moral, social, and political capital over the last approximately 2000 years.
	\end{itemize}

	\item D. M. Lydon-Staley, D. Zhou, A. S. Blevins, P. Zurn, and D. S. Bassett, “Hunters, busybodies, and the knowledge network building associated with curiosity,” 2019.
		\begin{itemize} 
		\item This paper provides empirical support for the kinesthetic curiosity model, finding convergent validity with psychometric approaches to measuring trait curiosity.
		\end{itemize}
		
	\item J. Gottlieb and P.-Y. Oudeyer, “Towards a neuroscience of active sampling and curiosity,” Nature Reviews Neuroscience, vol. 19, no. 12, pp. 758–770, 2018.
	\begin{itemize} 
		\item This review synthesizes the psychological, neuroscientific, and machine learning literature of attention and decision-making supporting the investigation of active information sampling, value, and search in a predictive processing framework for attention and decision-making. The authors emphasize the distinction between information sampling and search. To understand the full scope of curiosity-driven information search, the authors stress the importance of ecological complexity, little prior knowledge, and extended timescales, which are all characteristic of scientific research, education, and life-long learning.
	\end{itemize}
	
	\item G. M. Viswanathan, S. V. Buldyrev, S. Havlin, M. Da Luz, E. Raposo, and H. E. Stanley, “Optimizing the success of random searches,” Nature, vol. 401, no. 6756, pp. 911–914, 1999.
		\begin{itemize} 
		\item This paper develops the theory of optimal search using the statistical physics of L\'evy flight and finds empirical support in animal foraging data. 
		\end{itemize}
	
	\item I. Iacopini, S. c. v. Milojevic, and V. Latora, “Network dynamics of innovation processes,” Phys. Rev. Lett., vol. 120, p. 048301, Jan 2018.
		\begin{itemize} 
		\item This paper develops the theory of edge reinforcement network dynamics to model exploring novelties based on prior knowledge, and then applies it to the practice of scientific research.
		\end{itemize}
		
	\item J. Schmidhuber, “Driven by compression progress: A simple principle explains essential aspects of subjective beauty, novelty, surprise, interestingness, attention, curiosity, creativity, art, science, music, jokes,” in Workshop on anticipatory behavior in adaptive learning systems, pp. 48–76, Springer, 2008.
	    \begin{itemize} 
	    \item This textbook chapter develops the compression progress theory of curiosity and a predictive processing framework for machine learning and robotics. The author considers learning and creativity in light of compression progress theory.
	    \end{itemize}
	
	\item M. M. Garvert, R. J. Dolan, and T. E. Behrens (2017). A map of abstract relational knowledge in the human hippocampal–entorhinal cortex. Elife, 6, e17086.
	\begin{itemize} 
    	\item This paper provides evidence that hippocampal activity previously associated with spatial navigation also extends to the non-spatial navigation of abstract cognitive maps. Participants constructed cognitive maps without conscious awareness of the relationships among the map's discrete elements. The distance between elements in the map was defined as communicability, a graph theory measure discounting states separated by long distances. Communicability in network science closely resembles the successor representation in cognitive science.
	\end{itemize}
	
    \item K.L. Stachenfeld, M.M. Botvinick, and S.J. Gershman (2017). The hippocampus as a predictive map. Nature Neuroscience, 20(11), 1643.
    	\begin{itemize} 
    	\item This review develops a predictive coding theory of cognitive maps in the hippocampus. The hippocampus encodes compressed representations of possible future trajectories as the successor representation. The authors posit that place cells of the hippocampus encode the successor representation. Discrete grid modules and bottleneck states act as waypoints for optimal trajectories.
	\end{itemize}
	
	\item C. W. Lynn, L. Papadopoulos, A. E. Kahn, and D. S. Bassett, “Human information processing in complex networks,” Nature Physics (in press), 2020.
    	\begin{itemize} 
    	\item This paper studies what structural forms of ecologically relevant information support the communication and learning of sequences of information units. The authors address this question by developing a theory using statistical physics and information theory, and they apply it to empirical information networks in the domains of language, music, social relationships, internet websites, and scientific research citations. They find that highly informative and efficient forms of information networks have highly connected hubs and tightly linked modules. The model of temporal integration of learned structure used in this work can be derived separately from at least four other models: the temporal context model of memory, the successor representation in reinforcement learning, temporal difference learning, and the free energy principle of information theory.
    	\end{itemize}
    	
	\item M. Rosvall and C. T. Bergstrom (2008). Maps of random walks on complex networks reveal community structure. Proceedings of the National Academy of Sciences, 105(4), 1118-1123.
	\begin{itemize} 
    	\item This paper introduces the map algorithm that infers complex network structure from the information theoretic cost of encoding random walk dynamics.
	\end{itemize}

\end{itemize}

\newpage

\section{Acknowledgements}
We are thankful for the insightful feedback and comments from Dr. Linden Parkes and Jennifer Stiso. We gratefully acknowledge support from the John D. and Catherine T. MacArthur Foundation, the Alfred P. Sloan Foundation, the ISI Foundation, the Paul Allen Foundation, the Army Research Laboratory (No. W911NF-10-2-0022), the Army Research Office (Nos. Bassett-W911NF-14-1-0679, Grafton-W911NF-16-1-0474, and DCIST-W911NF-17-2-0181), the Office of Naval Research (ONR), the National Institute of Mental Health (Nos. 2-R01-DC-009209-11, R01-MH112847, R01-MH107235, and R21-M MH-106799), the National Institute of Child Health and Human Development (No. 1R01HD086888-01), National Institute of Neurological Disorders and Stroke (No. R01 NS099348), and the National Science Foundation (NSF) (Nos. DGE-1321851, BCS-1441502, BCS-1430087, NSF PHY-1554488, and BCS-1631550). The content is solely the responsibility of the authors and does not necessarily represent the official views of any of the funding agencies.

\newpage

\section{Citation Diversity Statement}

Recent work in neuroscience and other fields has identified a bias in citation practices such that papers from women and other minorities are under-cited relative to the number of such papers in the field \cite{Dworkin2020.01.03.894378}. Here we sought to proactively consider choosing references that reflect the diversity of the field in thought, form of contribution, gender, and other factors. We used automatic classification of gender based on the first names of the first and last authors \cite{Dworkin2020.01.03.894378}, with possible combinations including man/man, man/woman, woman/man, and woman/woman. Excluding self-citations to the first and last authors of our current paper, the references contain $52.8\%$ man/man, $12.4\%$ man/woman, $21.3\%$ woman/man, $13.5\%$ woman/woman, and $0\%$ unknown categorization. We look forward to future work that could help us to better understand how to support equitable practices in science.

\newpage
\bibliographystyle{ieeetr}
\bibliography{currOpinionsPsychCuriosity_v1.bib}

\end{document}